# Vibrational Strong Coupling of Ro-Vibrational States of Gaseous N$_2$O Molecules

*Akhila Kadyan and Jino George\**

Department of Chemical Sciences, Indian Institute of Science Education and Research (IISER) Mohali, Punjab-140306, India.

**Corresponding Author**

*jgeorge@iisermohali.ac.in

**ABSTRACT:** This letter demonstrates the formation of vibro-polaritonic states by strong coupling gaseous nitrous oxide molecules at room temperature. The ro-vibrational features of the molecule is utilized to understand the nature of the strong light-matter interaction in the gas phase. Mixing of the P and R-branch population gives three polaritonic branches, and the dispersion experiments are well supported by three coupled oscillator model. The best mixing of the ro-vibrational population occurs for the middle vibro-polaritonic state, which is formed at the place of the Q-branch (which is otherwise not allowed in the uncoupled system). Further, the Boltzmann distribution is also altered in the strong coupled system. High-resolution spectroscopic studies suggest the presence of a large uncoupled population along with the newly formed vibro-polaritonic states. The gas phase vibrational strong coupling on small molecules proposed here is one step closer to understanding the rate modification in polaritonic chemistry experiments.

Strong interactions between light and matter can result in the formation of hybrid states, cherishing both their properties.[1,2,3,4] These quasi-Bosonic states are first reported on atomic states interacting with a photon and are very well explained by the notion of cavity quantum electrodynamics.[5] Recently, there has been a surge in using the same concept in strong light-molecule interactions. However, the anharmonicity and the availability of rotational/vibrational/electronic states of the molecule make the system more difficult to study. This complexity offers a lot to explore with molecular systems. For example, both electronic and vibrational strong coupling are found to modify the chemical and physical properties of molecules. Recent experiments suggest that conductivity,[6,7,8,9,10] phase transition,[11] ferromagnetism,[12] energy transfer,[13,14,15,16] supramolecular assembly,[17] and crystallization[18], etc., can be modified by either coupling an electronic or vibrational transition to the confined electromagnetic field. The most important experiment in this direction is the modification of the chemical reaction rate reported in a few molecular systems by targeting a vibrational band associated with the reaction coordinate.[19,20,21,22,23,24,25,26] This may result in a reshuffling of the energy landscape or affect the reaction dynamics under strong coupling conditions.

There are many attempts, both in theory and experiment, to understand the mechanism of the process. Most of the experiments so far reported on polaritonic chemistry are either in liquid or solid phase. This makes the analysis very difficult as molecule-molecule or molecule-solvent interaction dominates in most of the cases. A few experimental and theoretical evidence suggest that the collective nature of the interaction is crucial to achieving a visible change in the reaction rates.[27,28,29,30,31,32] However, the involvement of strong coupling is questioned in many reported literature.[33,34] Very recent experiment suggests that the inter/intramolecular interactions and van der Waals forces can also be affected by coupling molecular transition to a cavity mode.[35] Solute-solvent interaction may be crucial in deciding the rate modification in such experiments.[36] In any condensed phase experiment, solute-solute interaction, such as self-dipolar interaction, can also be modified, which may even affect the transition probabilities.[37] The only way to study the effect of strong coupling on single molecules is to conduct an infinite dilution experiment or to do it in the gas phase. There are a few attempts in this direction to couple rotational states to a Fabry-Perot (FP) cavity and study the nature of polaritonic states.[38,39] The current situation demands both experimental and theoretical investigation of vibrational strong coupling in the gas phase.

Here, we chose a triatomic molecule- nitrous oxide ($N_2O$) and executed experiments in a specially prepared FP cavity configuration. The strongest vibrational state ($\nu_3$) of $N_2O$ is coupled to the cavity mode, which undergoes strong coupling at a reasonable concentration (2 bar pressure) in the gas phase. The evolution of vibro-polaritonic states while varying the pressure is also monitored by fixing the temperature (room temperature). Here, we also mapped the high-resolution ro-vibrational spectra of both the coupled and uncoupled states. P and R-branches of the ro-vibrational states get mixed up in the cavity, forming upper, middle, and lower polaritonic states. Further, we attempt to understand the nature of ro-vibrational state mixing in the strongly coupled gaseous molecules.

$N_2O$, being a linear triatomic molecule, exhibits four normal modes of vibration. Symmetric stretching ($\nu_1$) at 1285cm$^{-1}$, asymmetric stretching ($\nu_3$) at 2224cm$^{-1}$, two degenerate bending modes ($\nu_2$) at 589cm$^{-1}$.[40] Unlike $CO_2$, all of the fundamental vibrational bands of $N_2O$ are both IR and Raman active as it is a non-centrosymmetric system. In the gas phase, molecules will have rotational transitions simultaneously with vibrational transitions. This results in additional spectral features in the form of rotational transitions superimposed upon the vibrational bands, generally known as the ro-vibrational spectrum. Here, for the triatomic $N_2O$ molecules, the rotational transitions are classified as P, Q, and R-branches. The P and R-branches of asymmetric stretching undergo VSC, thereby forming upper (UP), middle (MP), and lower (LP) vibro-polaritonic states, as shown in Figure 1. Commercially available cells are typically centimeters long and are commonly used to record the IR spectra of gaseous molecules. However, the required cell pathlength for executing the VSC should be within a few micrometers. Therefore, we used a demountable microfluidic cavity (in a flow-cell condition) and stabilized the mirrors by placing them inside an isolation chamber to perform VSC of gaseous $N_2O$ molecules (figure 2). Further, to prevent the IR transparent $BaF_2$ substrates from breaking, we slowly increased the pressure inside the cavity, and the IR spectra are recorded with a continuous purging of $N_2O$ gas. Additionally, a pressure gauge is mounted at the $N_2O$ gas inlet, and the spectrum is recorded at various inlet pressures. Further, spectra are recorded at both low (2 cm$^{-1}$) and high (0.2 cm$^{-1}$) resolutions to get an insight into ro-vibrational characteristics (SI section 2).

FP cavities are prepared by coating 6 nm gold mirror to one side of the $BaF_2$ substrates and separating them using a 25 μm mylar spacer (Figure 2). The flow-cell FP cavity is placed inside the VT cell holder unit, and all the experiments are performed by using the same configuration at room temperature. 11$^{th}$ cavity mode is set to couple the asymmetric stretching band of $N_2O$ molecules. Further, the position of the cavity mode is tuned to become ON-resonance at three distinct positions: P-branch maximum, R-branch maximum, and vibrational band center (Q-branch position). The newly formed vibro-polaritonic states (UP, MP, and LP) disperse as the cavity mode is mixed with P-and R-branch populations. When the cavity mode is at ON-resonance with the P-branch population, the UP is coupled with the R-branch population, resulting in a total Rabi splitting energy of 41 cm$^{-1}$. The average full-width half maximum (FWHM) of the asymmetric band (49 cm$^{-1}$) and empty cavity mode (22 cm$^{-1}$) is 35cm$^{-1}$, which satisfies the minimum criteria for strong coupling. Moreover, the Rabi splitting energy for individual coupling (measured by TMM) of P and R-branches separately are 24 and 23 cm$^{-1}$, respectively. The estimated FWHM for bare P and R-branches are 27 and 19 cm$^{-1}$, respectively (more details are given in section 5 of SI). Tuning the cavity mode at different positions is represented in Figure 3b. There is a clear distribution of density of states as one tune across different branches. To get additional insight into the origin of three vibro-polaritonic states, Hopfield coefficients are calculated, which provide the extent of the R-branch, P-branch, and photonic content. This is achieved by constructing a $3X3$ matrix as follows:

$$\begin{bmatrix} E_C & (\hbar\Omega_R)_P/2 & (\hbar\Omega_R)_R/2 \\ (\hbar\Omega_R)_P/2 & E_P & 0 \\ (\hbar\Omega_R)_R/2 & 0 & E_R \end{bmatrix} \begin{bmatrix} \alpha \\ \beta \\ \gamma \end{bmatrix} = E_P \begin{bmatrix} \alpha \\ \beta \\ \gamma \end{bmatrix}$$

where, $E_C$, $E_P$, and $E_R$ is the energy of 11$^{th}$ cavity mode, P-branch maximum, and R-branch maximum, respectively. $(\hbar\Omega_R)_P$ and $(\hbar\Omega_R)_R$ represents the corresponding Rabi splitting energy when 11$^{th}$ cavity mode is at ON-resonance with respect to P and R-branch maxima. $E_P$ represents the energy of the vibro-polaritonic state, which can be either UP, MP, or LP. $|\alpha|^2$, $|\beta|^2$ and $|\gamma|^2$ are the Hopfield coefficients corresponding to photon, P-branch, and R-branch fractions, respectively. The angle-dependent dispersion of vibro-polaritonic states is calculated in TMM, and it complements the experimental observation, as shown in Figure 3a. A plot of the Hopfield coefficients of the MP is given in Figure 3c. Further, the Hopfield parameters for UP and LP are given in the SI (figure S6). Analyzing the photon, P-branch, and R-branch fractions shows that a clear mixing of the states occurs at 6° angle of incidence. Here, both P-branch and R-branch fractions are equally contributing to the MP and are uniquely originating from different angular momentum conserved states. Hence, the MP may inherit both the properties of the P and R-branches but still obey the conservation law.

Further, high-resolution (HR) IR experiments (resolution of 0.20 cm$^{-1}$) are conducted to gain deep insight into the mixing of ro-vibrational states of N$_2$O molecules. The solid blue line in Figure 4 shows the HR non-cavity asymmetric stretching band of N$_2$O at an inlet pressure of 2 bar. The asymmetric stretching bands show P and R-branches maxima at 2212 cm$^{-1}$ and 2237 cm$^{-1}$, respectively. The inset plot in Figure 4 shows a zoomed HR spectra of the first few rotational lines of P and R-branches. By carefully looking into the position of individual rotational lines and performing simple mathematical calculations, the rotational constant of the ground vibrational state (B$_0$) is found to be 0.42 cm$^{-1}$, which is in good agreement with the literature report.[40] The calculated $J_{max}$ is found to be 15, meaning the rotational transition from $J_{15\to16}$ of the ground vibrational state has the highest population in both P and R-branches. Very interestingly, the cavity HR spectra show three vibro-polaritonic branches when cavity mode is positioned at ON-resonance with the P-branch maximum (solid red line in Figure 4). The P and R-branches maxima are now reshuffled across the newly formed vibro-polaritonic states. Additionally, the rotational fingers are also visible on top of the three vibro-polaritonic states. However, there appears to be no change in the position of the individual rotational lines under VSC conditions when compared with non-cavity. This indicates that rotational constants remain intact under VSC conditions. The above observation makes sense that the rotation dynamics are approximately three orders lower than vibrational dynamics. Hence, the Rabi interaction may not be influencing the rotational quantum states. In other words, the symmetry of the molecular dipole is still intact even if the vibrational envelope is coupled to the cavity mode. Most importantly, the formation of MP at the position of the Q-branch is clear; otherwise, it is not allowed in non-cavity conditions. This is due to the reshuffling of the population density under VSC condition. i.e., the concept of Maxwell-Boltzmann distribution and the $J_{max}$ population under the cavity coupled situation no longer valid.

Another possibility of seeing the rotational fingers in a strongly coupled system is the availability of uncoupled N₂O molecules. A large fraction of N₂O molecules is not coupled to the cavity mode and rides on top of the vibro-polaritonic states. At the moment, we cannot predict the percentage fraction of the uncoupled population, and the only way to distinguish them is to enter into very high coupling strength by ramping up the pressure of the cavity system. This may allow us to distinguish the vibro-polaritonic features from the uncoupled population. The FP cavity configuration presented here can go a maximum of 2 bars; afterward, the mirrors are unstable to study experimentally. These experiments are challenging, and we are currently making more robust gas phase cells to address these issues. Nevertheless, mixing P- and R-branch populations is very interesting as the transitions originate from different angular momentum selection rules. Another interesting aspect of the current work is that the rotational states may scramble up as the ro-vibrational transition is coupled to a cavity mode, thereby affecting the rotation of the molecule with respect to the bond axis. If this scenario is true, the vacuum field can arrest the rotation, and hence, the effective change in the rotational dipole moment becomes zero. Similar scrambling experiments are already known in the literature, which normally happen at very high power of infrared laser excitation. The classical nature of the rotational states disappears as the vibrational dipoles are forced to align with respect to the electromagnetic field. Our intention was to test the symmetry breaking under VSC condition. Current experimental challenges may not allow us to comment on these aspects.

In conclusion, we have successfully demonstrated the formation of vibro-polaritonic states in the gas phase at room temperature. The availability of ro-vibrational states gives more information on the nature of strong light-matter interaction in molecular systems. Here, the mixing of P- and R-branch populations is yet another way of studying three coupled oscillator model, but originating from the same vibrational transition. These findings on small molecules, especially triatomic systems, open new avenues to study the origin of chemical reaction modification under VSC, such as unimolecular decomposition reactions.

## ACKNOWLEDGMENT

A. K. thank IISER Mohali for the research fellowship. J. G. thank IISER Mohali for the seed grant and the infrastructure facilities. The above work is supported by funding from MoE, India, Scheme for Transformational and Advanced Research in Sciences (**MoE-STARS/STARS-1/175**).

## REFERENCES

(1) Ebbesen, T. W. Hybrid Light-Matter States in a Molecular and Material Science Perspective. *Acc. Chem. Res.* **2016**, *49* (11), 2403–2412. https://doi.org/10.1021/acs.accounts.6b00295.

(2) Hertzog, M.; Wang, M.; Mony, J.; Börjesson, K. Strong Light-Matter Interactions: A New Direction within Chemistry. *Chem. Soc. Rev.,* **2019**, *48 (3), 937-961*. https://doi.org/10.1039/C8CS00193F.

(3) Hirai, K.; Hutchison, J. A.; Uji-I, H. Molecular Chemistry in Cavity Strong Coupling. *Chem. Rev.* **2023**, *123* (13), 8099–8126. https://doi.org/10.1021/acs.chemrev.2c00748.

(4) Mandal, A.; Taylor, M. A. D.; Weight, B. M.; Koessler, E. R.; Li, X.; Huo, P. Theoretical Advances in Polariton Chemistry and Molecular Cavity Quantum Electrodynamics. *Chem. Rev.* **2023**, *123* (16), 9786–9879. https://doi.org/10.1021/acs.chemrev.2c00855.

(5) Haroche, S.; Kleppner, D. Cavity Quantum Electrodynamics. *Phys. Today* **1989**, *42* (1), 24–30. https://doi.org/10.1063/1.881201.

(6) Orgiu, E.; George, J.; Hutchison, J. A.; Devaux, E.; Dayen, J. F.; Doudin, B.; Stellacci, F.; Genet, C.; Schachenmayer, J.; Genes, C.; Pupillo, G.; Samorì, P.; Ebbesen, T. W. Conductivity in organic semiconductors hybridized with the Vacuum Field. *Nat. Mater.* **2015**, *14* (September), 1123–1130. https://doi.org/10.1038/NMAT4392.

(7) Hagenmüller, D.; Schachenmayer, J.; Schütz, S.; Genes, C.; Pupillo, G. Cavity-Enhanced Transport of Charge. *Phys. Rev. Lett.* **2017**, *119* (22), 223601. https://doi.org/10.1103/PhysRevLett.119.223601.

(8) Kaur, K.; Johns, B.; Bhatt, P.; George, J. Controlling Electron Mobility of Strongly Coupled Organic Semiconductors in Mirrorless Cavities. *Adv. Funct. Mater.* **2023**, *33* (47), 2306058. https://doi.org/10.1002/adfm.202306058.

(9) Fukushima, T.; Yoshimitsu, S.; Murakoshi, K. Inherent Promotion of Ionic Conductivity via Collective Vibrational Strong Coupling of Water with the Vacuum Electromagnetic Field. *J. Am. Chem. Soc.* **2022**, *144*, 12177−12183. https://doi.org/10.1021/jacs.2c02991.

(10) Kumar, S.; Biswas, S.; Rashid, U.; Mony, K. S.; Vergauwe, R. M. A.; Kaliginedi, V.; Thomas, A. Extraordinary Electrical Conductance of Non-Conducting Polymers Under Vibrational Strong Coupling. *ChemRxiv* **2023**, 1–14.

(11) Wang, S.; Mika, A.; Hutchison, J. A.; Genet, C.; Jouaiti, A.; Hosseini, M. W.; Ebbesen, T. W. Phase Transition of a Perovskite Strongly Coupled to the Vacuum Field. *Nanoscale* **2014**, *6* (13), 7243–7248. https://doi.org/10.1039/c4nr01971g.

(12) Thomas, A.; Devaux, E.; Nagarajan, K.; Rogez, G.; Seidel, M.; Richard, F.; Genet, C.; Drillon, M.; Ebbesen, T. W. Large Enhancement of Ferromagnetism under a Collective

Strong Coupling of YBCO Nanoparticles. *Nano Lett.* **2021**, *21* (10), 4365–4370. https://doi.org/10.1021/acs.nanolett.1c00973.

(13) Zhong, X.; Chervy, T.; Wang, S.; George, J.; Thomas, A.; Hutchison, J. A.; Devaux, E.; Genet, C.; Ebbesen, T. W. Non-Radiative Energy Transfer Mediated by Hybrid Light-Matter States. *Angew. Chemie* **2016**, *128* (21), 6310–6314. https://doi.org/10.1002/ange.201600428.

(14) Zhong, X.; Chervy, T.; Zhang, L.; Thomas, A.; George, J.; Genet, C.; Hutchison, J. A.; Ebbesen, T. W. Energy Transfer between Spatially Separated Entangled Molecules. *Angew. Chemie - Int. Ed.* **2017**, *56* (31), 9034–9038. https://doi.org/10.1002/anie.201703539.

(15) Wang, M.; Hertzog, M.; Börjesson, K. Polariton-Assisted Excitation Energy Channeling in Organic Heterojunctions. *Nat. Commun.* **2021**, *12* (1), 1874. https://doi.org/10.1038/s41467-021-22183-3.

(16) Bhatt, P.; Dutta, J.; Kaur, K.; George, J. Long-Range Energy Transfer in Strongly Coupled Donor − Acceptor Phototransistors. **2023**. https://doi.org/10.1021/acs.nanolett.3c00867.

(17) Joseph, K.; Kushida, S.; Smarsly, E.; Ihiawakrim, D.; Thomas, A.; Paravicini-bagliani, G. L.; Nagarajan, K.; Vergauwe, R.; Devaux, E.; Ersen, O.; Bunz, U. H. F.; Ebbesen, T. W. Supramolecular Chemistry Hot Paper Supramolecular Assembly of Conjugated Polymers under Vibrational Strong Coupling. *Angew. Chemie - Int. Ed.* **2021**, *60*, 19665–19670. https://doi.org/10.1002/anie.202105840.

(18) Hirai, K.; Ishikawa, H.; Chervy, T.; Hutchison, J. A.; Uji-i, H. Selective Crystallization via Vibrational Strong Coupling. *Chem. Sci.* **2021**, *12* (36), 11986–11994. https://doi.org/10.1039/d1sc03706d.

(19) Thomas, A.; George, J.; Shalabney, A.; Dryzhakov, M.; Varma, S. J.; Moran, J.; Chervy,

T.; Zhong, X.; Devaux, E.; Genet, C.; Hutchison, J. A.; Ebbesen, T. W. Ground-State Chemical Reactivity under Vibrational Coupling to the Vacuum Electromagnetic Field. *Angew. Chemie* **2016**, *128* (38), 11634–11638. https://doi.org/10.1002/ange.201605504.

(20) Lather, J.; Bhatt, P.; Thomas, A.; Ebbesen, T. W.; George, J. Cavity Catalysis by Cooperative Vibrational Strong Coupling of Reactant and Solvent Molecules. *Angew. Chemie* **2019**, *131* (31), 10745–10748. https://doi.org/10.1002/ange.201905407.

(21) Hirai, Kenji, Rie Takeda, James A. Hutchison, and H. U. Modulation of Prins Cyclization by Vibrational Strong Coupling. *Angew. Chemie Int. Ed.* **2020**, *132* (13), 5370–5373. https://doi.org/10.1002/anie.201915632.

(22) Ahn, W.; Triana, J. F.; Recabal, F.; Herrera, F.; Simpkins, B. S. Modification of Ground-State Chemical Reactivity via Light – Matter Coherence in Infrared Cavities. *Science* **2023**, *380* (6650), 1165–1168. https://doi.org/10.1126/science.ade7147.

(23) Vergauwe, R. M. A.; Thomas, A.; Nagarajan, K.; Shalabney, A.; George, J.; Chervy, T.; Seidel, M.; Torbeev, V.; Ebbesen, T. W. Modification of Enzyme Activity by Vibrational Strong Coupling of Water. *Angew. Chemie Int. Ed.* **2019**, *58*, 15324–15328. https://doi.org/10.1002/anie.201908876.

(24) Lather, J.; George, J. Improving Enzyme Catalytic Efficiency by Co-Operative Vibrational Strong Coupling of Water. *J. Phys. Chem. Lett.* **2021**, *12* (1), 379–384. https://doi.org/10.1021/acs.jpclett.0c03003.

(25) Gu, K.; Si, Q.; Li, N.; Gao, F.; Wang, L.; Zhang, F. Regulation of Recombinase Polymerase Amplification by Vibrational Strong Coupling of Water. *ACS Photonics* **2023**, *10* (5), 1633–1637. https://doi.org/10.1021/acsphotonics.3c00243.

(26) Bai, J.; Wang, Z.; Zhong, C.; Hou, S.; Lian, J.; Si, Q. Vibrational Coupling with OH

Stretching Increases Catalytic Efficiency of Sucrase in Fabrye Perot Microcavity. *Biochem. Biophys. Res. Commun.* **2023**, *652*, 31–34. https://doi.org/10.1016/j.bbrc.2023.02.025.

(27) Wang, D. S.; Yelin, S. F. A Roadmap Toward the Theory of Vibrational Polariton Chemistry. *ACS Photonics* **2021**, *8* (10), 2818–2826. https://doi.org/10.1021/acsphotonics.1c01028.

(28) Schäfer, C.; Ruggenthaler, M.; Rokaj, V.; Rubio, A. Relevance of the Quadratic Diamagnetic and Self-Polarization Terms in Cavity Quantum Electrodynamics. *ACS Photonics* **2020**, *7* (4), 975–990. https://doi.org/10.1021/acsphotonics.9b01649.

(29) Schäfer, C.; Flick, J.; Ronca, E.; Narang, P.; Rubio, A. Shining Light on the Microscopic Resonant Mechanism Responsible for Cavity-Mediated Chemical Reactivity. *Nat. Commun.* **2022**, *13* (1), 1–9. https://doi.org/10.1038/s41467-022-35363-6.

(30) Flick, J.; Rivera, N.; Narang, P. Strong Light-Matter Coupling in Quantum Chemistry and Quantum Photonics. *Nanophotonics*. De Gruyter 2018, pp 1479–1501. https://doi.org/10.1515/nanoph-2018-0067.

(31) Hu, D.; Ying, W.; Huo, P. Resonance Enhancement of Vibrational Polariton Chemistry Obtained from the Mixed Quantum-Classical Dynamics Simulations. *J. Phys. Chem. Lett.* **2023**, *14* (49), 11208−11216. https://doi.org/10.1021/acs.jpclett.3c02985.

(32) Li, X.; Mandal, A.; Huo, P. Cavity Frequency-Dependent Theory for Vibrational Polariton Chemistry. *Nat. Commun.* **2021**, *12* (1), 1–9. https://doi.org/10.1038/s41467-021-21610-9.

(33) Imperatore, M. V.; Asbury, J. B.; Giebink, N. C. Reproducibility of Cavity-Enhanced Chemical Reaction Rates in the Vibrational Strong Coupling Regime. *J. Chem. Phys.* **2021**, *154* (19), 191103. https://doi.org/10.1063/5.0046307.

(34) Wiesehan, G. D.; Xiong, W. Negligible Rate Enhancement from Reported Cooperative

Vibrational Strong Coupling Catalysis. *J. Chem. Phys.* **2021**, *155* (24), 241103. https://doi.org/10.1063/5.0077549.

(35) Patrahau, B.; Piejko, M.; Mayer, R.; Antheaume, C.; Sangchai, T.; Ragazzon, G.; Jayachandran, A.; Devaux, E.; Genet, C.; Moran, J.; Ebbesen, T. W. Direct Observation of Polaritonic Chemistry by Nuclear Magnetic Resonance Spectroscopy. *ChemRxiv* **2023**.

(36) Piejko, M.; Patrahau, B.; Joseph, K.; Muller, C.; Devaux, E.; Ebbesen, T. W.; Moran, J. Solvent Polarity under Vibrational Strong Coupling. *J. Am. Chem. Soc.* **2023**, *145* (24), 13215–13222. https://doi.org/10.1021/jacs.3c02260.

(37) Kadyan, A.; Shaji, A.; George, J. Boosting Self-Interaction of Molecular Vibrations under Ultrastrong Coupling Condition. *J. Phys. Chem. Lett.* **2021**, *12* (17), 4313–4318. https://doi.org/10.1021/acs.jpclett.1c00552.

(38) Wright, A. D.; Nelson, J. C.; Weichman, M. L. Rovibrational Polaritons in Gas-Phase Methane. *J. Am. Chem. Soc.* **2023**, *145* (10), 5982–5987. https://doi.org/10.1021/jacs.3c00126.

(39) Wright, A. D.; Nelson, J. C.; Weichman, M. L. A Versatile Platform for Gas-Phase Molecular Polaritonics. *J. Chem. Phys.* **2023**, *159* (16), 164202. https://doi.org/10.1063/5.0170326.

(40) Plyler, E. K.; Tidwell, E. D.; Maki, A. G. Infrared Absorption Spectrum of Nitrous Oxide ($N_2O$) from 1830 $cm^{-1}$ to 2270 $cm^{-1}$. *J. Res. Natl. Bur. Stand. Sect. A Phys. Chem.* **1964**, *68A* (1), 79. https://doi.org/10.6028/jres.068a.006.

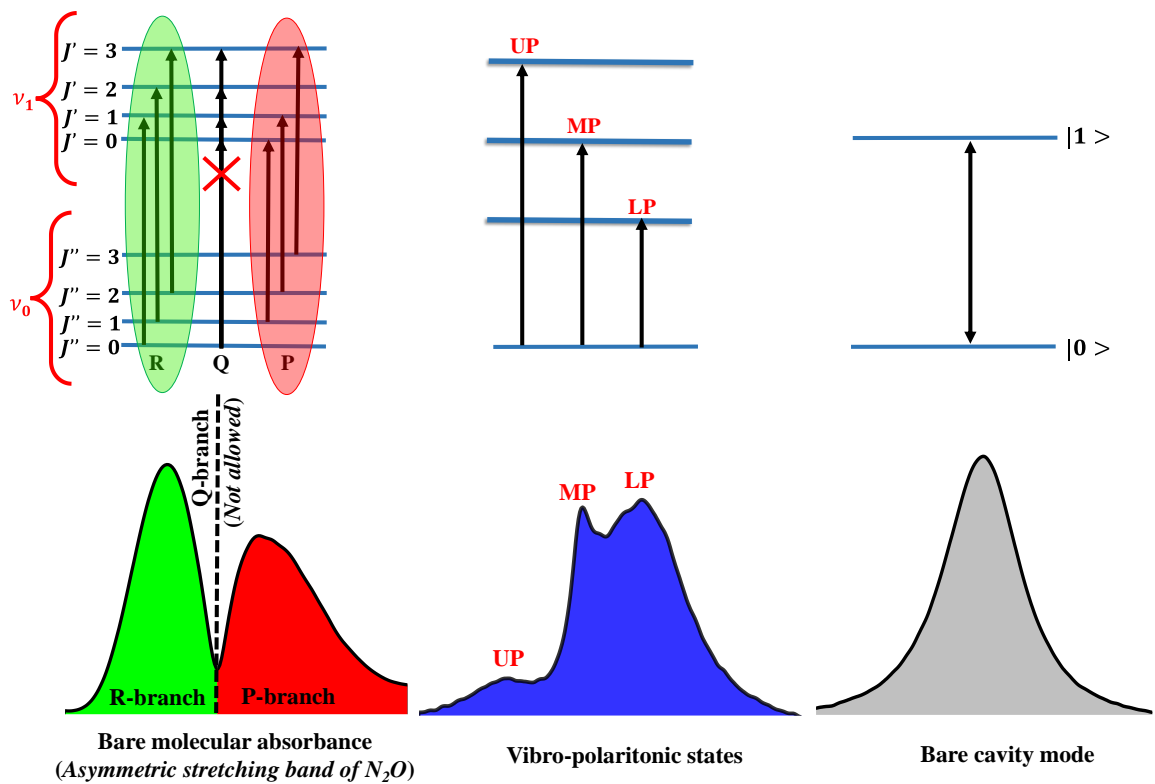

**Figure 1.** Schematic representation of VSC between ro-vibrational asymmetric stretching band of N$_2$O and an FP cavity mode shows the formation of ro-vibro-polaritonic states (UP, MP, and LP).

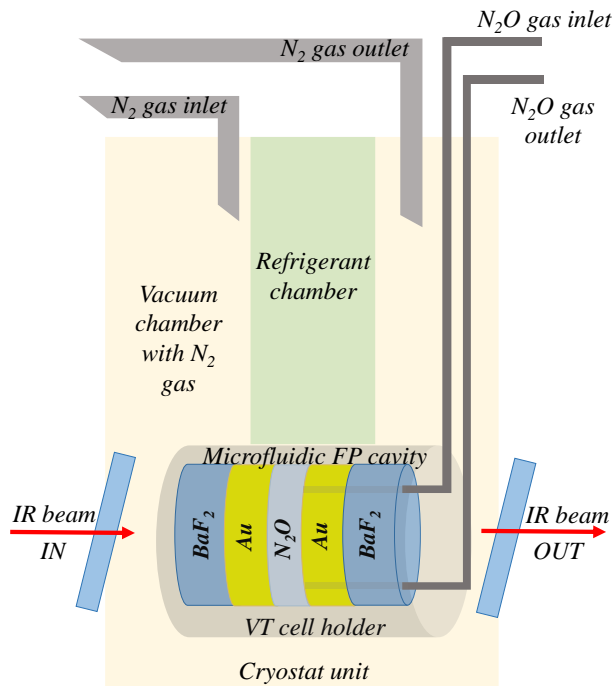

**Figure 2.** Drawing of the experimental setup for measuring ro-vibrational gas phase vibrational strong coupling of N$_2$O molecules.

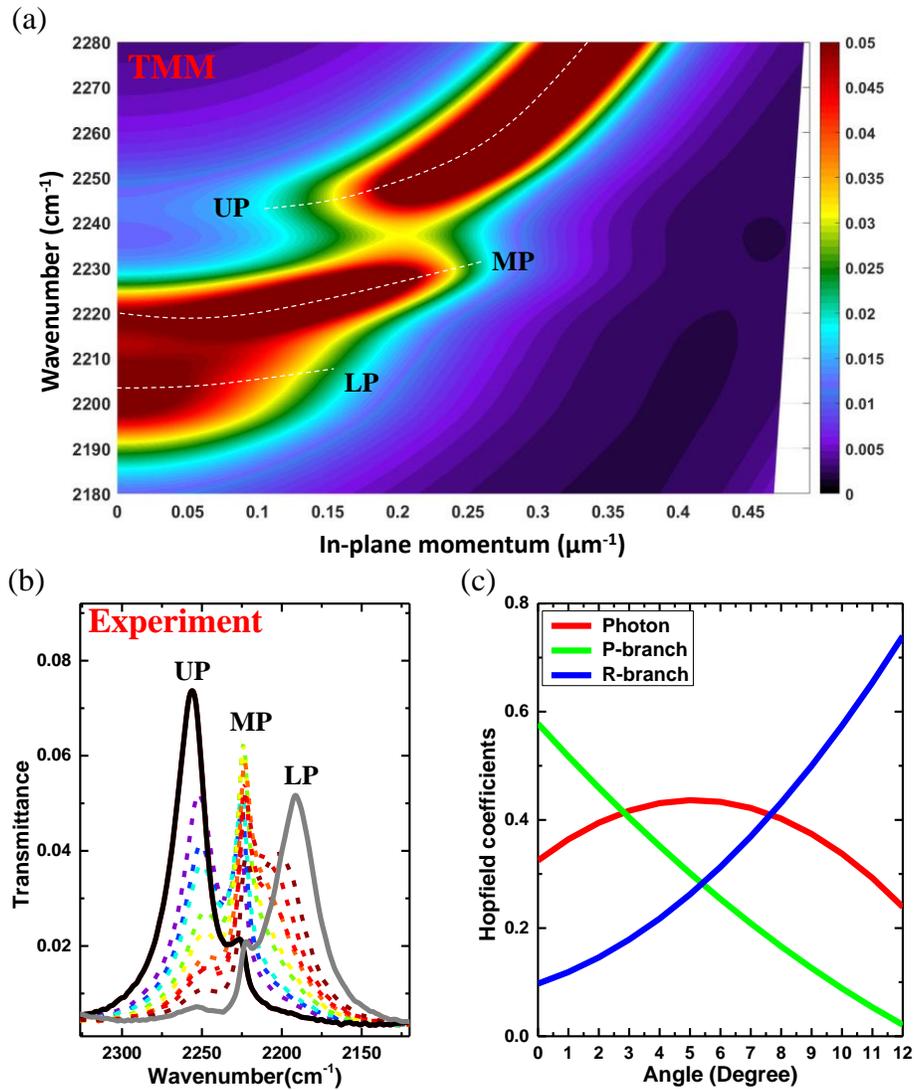

**Figure 3.** Dispersion of vibro-polaritonic states (UP, MP, and LP) obtained by coupling $N_2O$ molecules, **(a)** TMM simulation, and **(b)** experiments. **(c)** Hopfield coefficients correspond to the MP with photon, P-branch, and R-branch fractions.

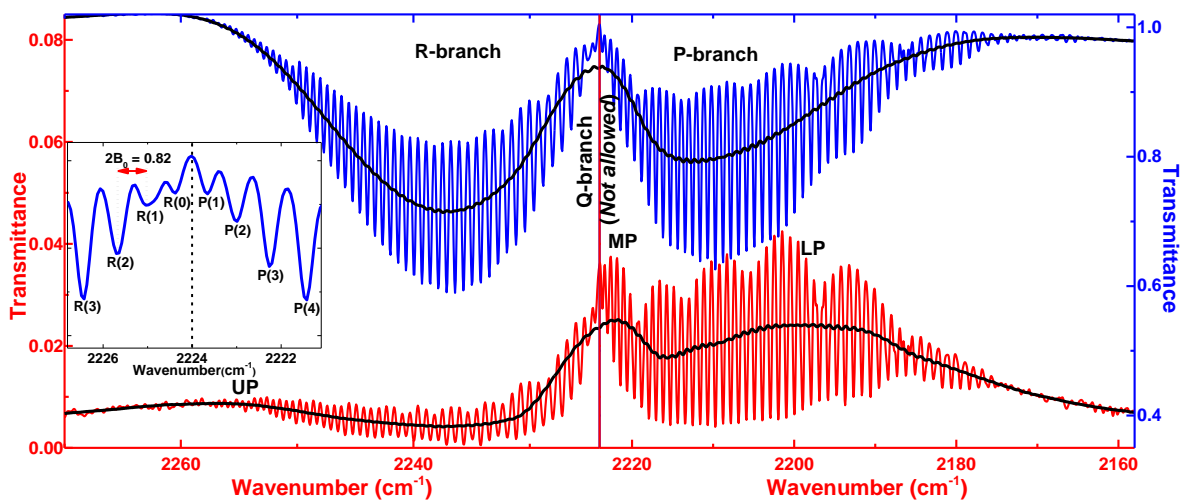

**Figure 4.** High-resolution IR spectra of the non-cavity (blue) and cavity (red) ro-vibrational states of $N_2O$ asymmetric stretching band. The black solid traces are the averaged spectral line-shapes. The inset plot shows the individual rotational lines in P and R-branches with a separation of $2B_0$ as 0.82 cm$^{-1}$.

# Supporting Information:

1. **Materials and methods**

The nitrous oxide gas is purchased from Sigma Gases India Pvt. Ltd. The demountable flow cell, $BaF_2$ substrates, mylar spacers, and cryostat unit used for the fabrication of flow-cell gas phase FP cavity is purchased from Specac Ltd., UK. The Fourier Transform Infrared (FT-IR) interferometer (© Bruker model INVENIO) is used for recording the IR transmission spectrum $N_2O$ molecules in both cavity and non-cavity.

2. **Non-cavity IR spectrum of $N_2O$ gas**

Firstly, the non-cavity IR transmission spectrum of $N_2O$ molecules is recorded in both low resolution (2 cm$^{-1}$) at different inlet pressures ranging from 0.25 to 2.00 bar (figure S1). It shows asymmetric stretching and symmetric stretching bands at 2224 and 1285 cm$^{-1}$, respectively. Both the asymmetric and symmetric stretching bands are accompanied by P and R branches. The bending mode is not observed since $BaF_2$ substrates are not IR transparent below 1000 cm$^{-1}$.

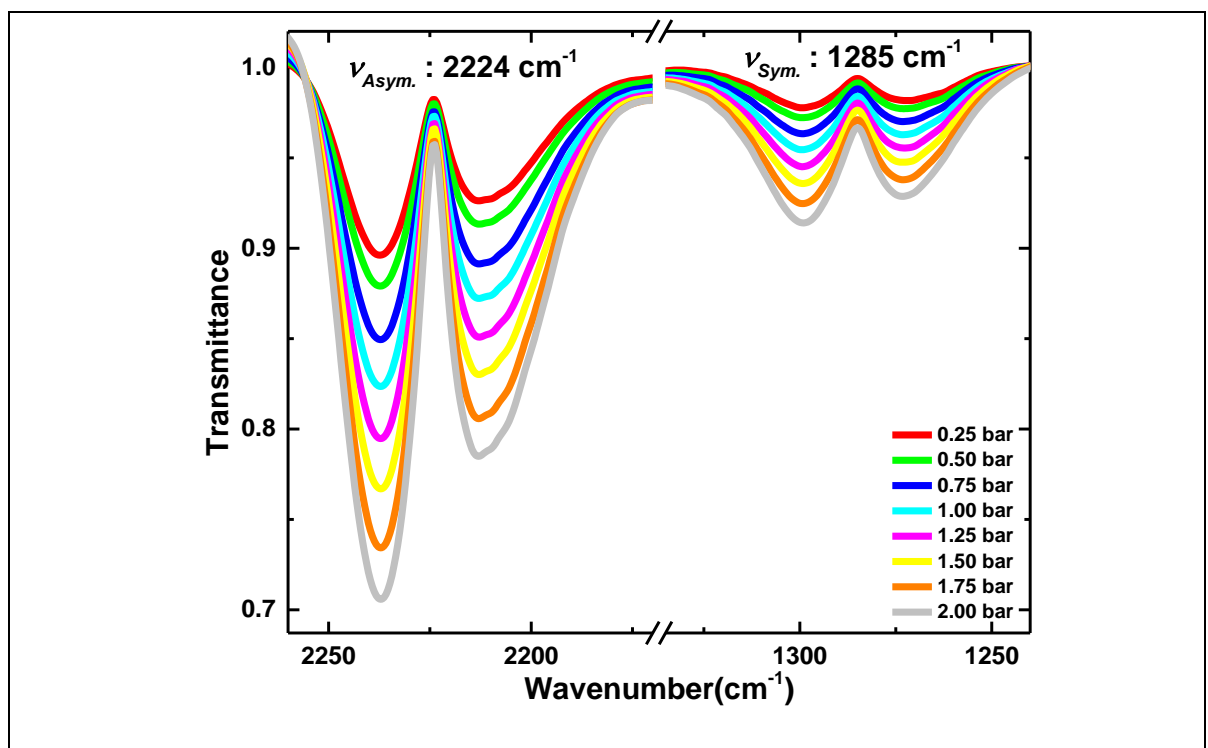

**Figure S1.** Experimentally obtained non-cavity IR transmission spectrum of $N_2O$ at different inlet pressures.

Similarly, the high-resolution (0.2 cm$^{-1}$) IR spectrum of N$_2$O gas is recorded which shows the individual rotational lines accompanying the vibrational bands (Figure S2).

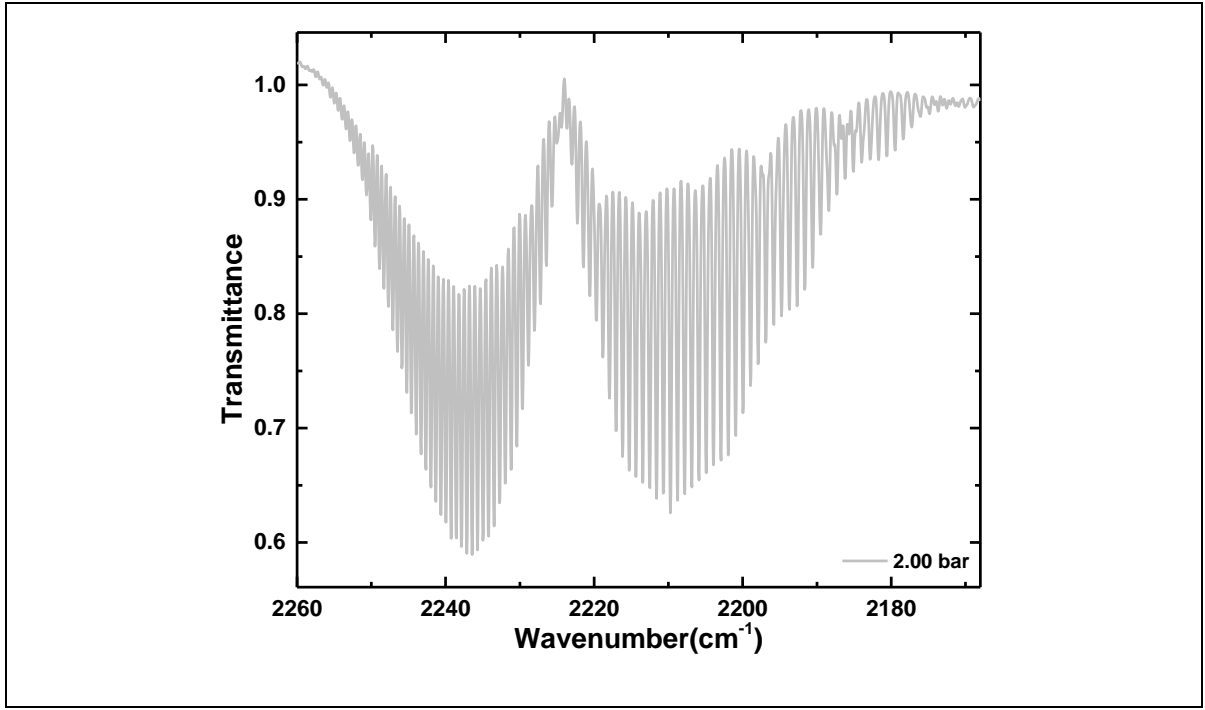

**Figure S2.** Experimentally obtained high-resolution non-cavity IR transmission spectrum showing asymmetric stretching band of N$_2$O with different rotational lines at an inlet pressure of 2 bar.

3. **Rotational constant calculation**

The energies of individual rotational lines in R and P branches are given by the following expressions:

$$R(J) = \bar{\omega}_0 + (J + 1)(B_1 + B_0) + (J + 1)^2(B_1 - B_0)$$

$$P(J) = \bar{\omega}_0 - J(B_1 + B_0) + J^2(B_1 - B_0)$$

By performing simple mathematics on the above equations, one can get

$$R(J - 1) - P(J + 1) = 2(2J + 1)B_0$$

From the above equation, by plotting $R(J - 1) - P(J + 1)$ as a function $2(2J + 1)$ yields a straight-line curve with a slope of $B_0$ as 0.41615 (Figure S3).

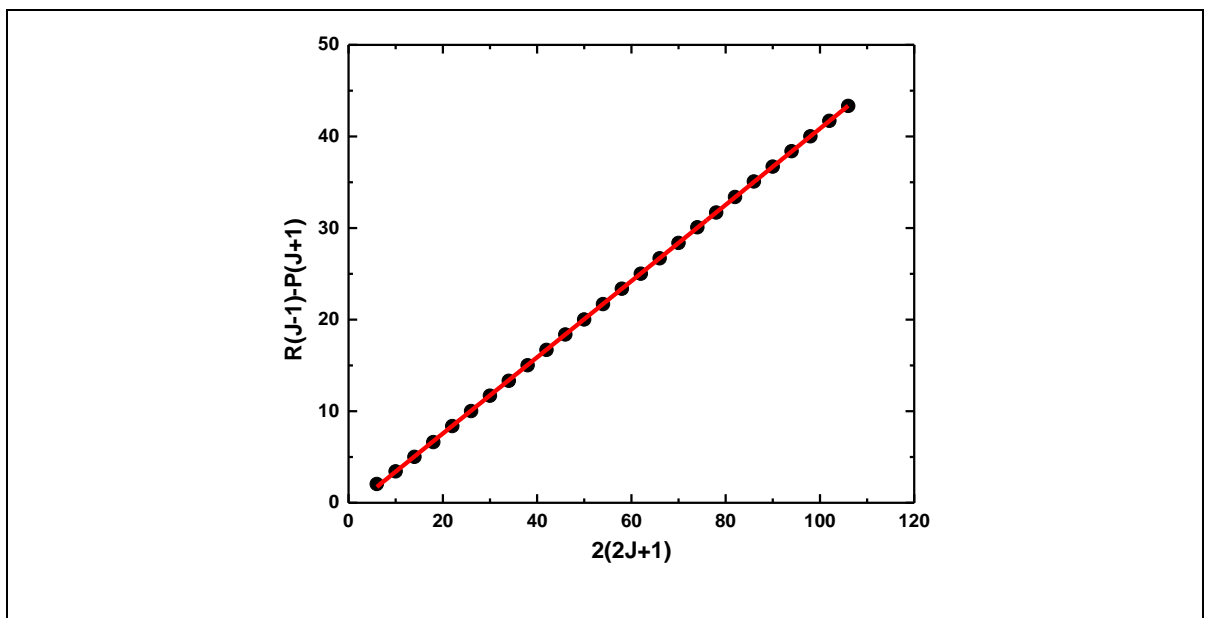

**Figure S3.** Plot of $R(J-1)-P(J+1)$ and $R(J)-P(J)$ versus $2(2J+1)$ with slopes equivalent to $B_0$.

## 4. Transfer matrix methods (TMM) fitting

A multi-Lorentzian function in TMM simulations is first used for fitting the low-resolution IR spectrum of asymmetric stretching band of $N_2O$ molecules in non-cavity condition (Figure S4(a)). Then TMM is used for calculating the Rabi splitting energy for VSC of asymmetric stretching band at three different positions (P-branch maximum, R-branch maximum and vibrational band centre). The TMM calculated Rabi splitting energy is found to be in good agreement with the experimentally obtained value (Figure S4).

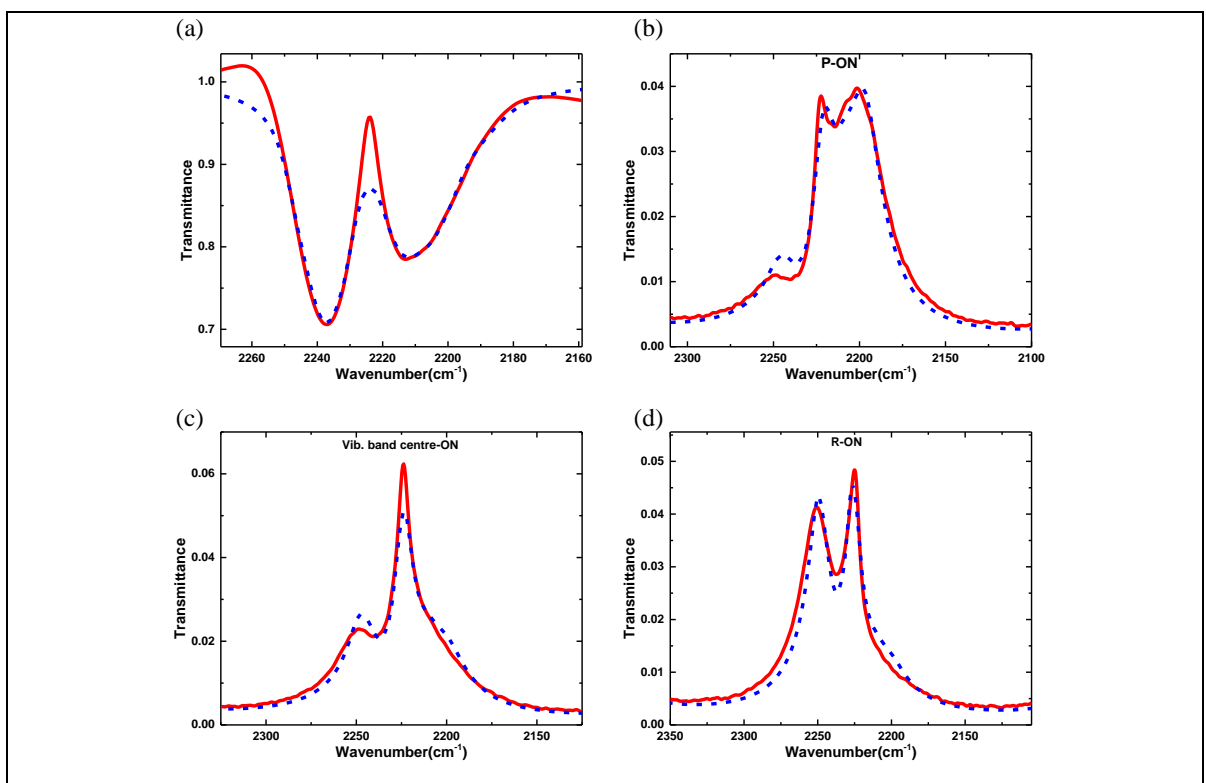

**Figure S4. (a)** TMM fitting of non-cavity IR asymmetric stretching band of $N_2O$. TMM fitting of VSC of asymmetric stretching band of $N_2O$ inside cavity with cavity mode ON resonance with **(b)** P-branch maximum, **(c)** vibrational band centre, and **(d)** R-branch maximum (Experiment: solid red line; TMM: dashed blue line)

## 5. TMM simulation for individual coupling of P and R branches

TMM simulations are performed to get information about the individual Rabi splitting energy of the P or R branch (Figure S5). The TMM simulations suggest that the Rabi splitting energy when cavity mode is coupled with P-branch and R-branch individually, is 24 cm$^{-1}$ and 23 cm$^{-1}$, respectively. The average of FWHM of the R-branch (19 cm$^{-1}$) and empty cavity mode (22 cm$^{-1}$) is 21 cm$^{-1}$. The average of FWHM of the P-branch (27 cm$^{-1}$) and empty cavity mode (22 cm$^{-1}$) is 25 cm$^{-1}$.

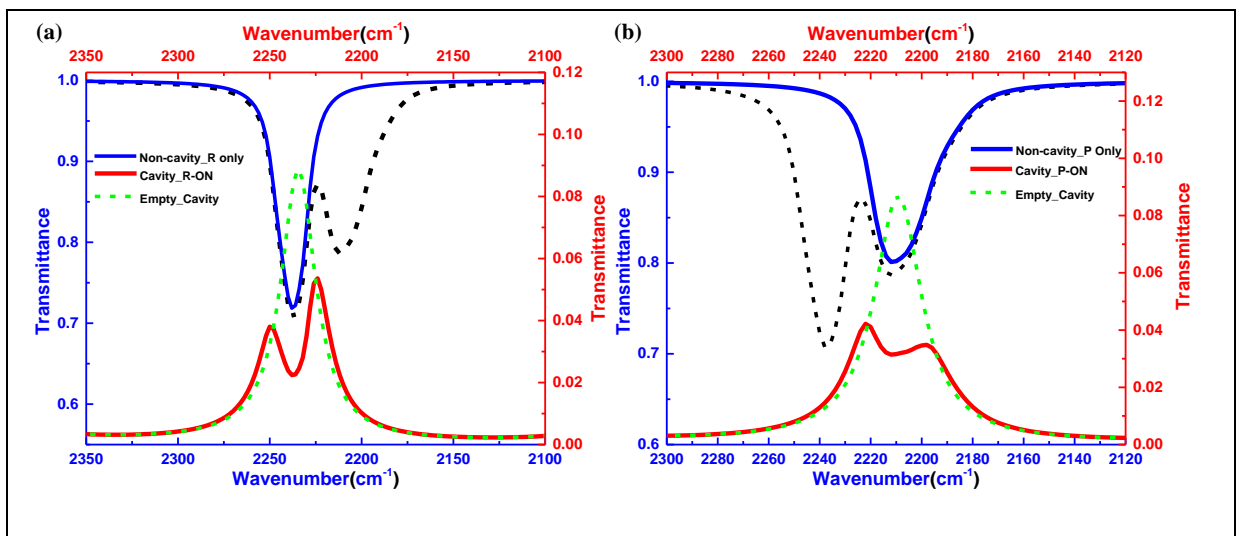

**Figure S5.** TMM calculated Rabi splitting energy for individual coupling of **(a)** R-branch, and **(b)** P-branch of asymmetric stretching band of N2O molecules. (solid blue line: non-cavity TMM fitting of induvial R or P branch; solid red line: TMM predicted polaritonic states formation for individual coupling of R or P branch; dotted green line: empty cavity mode; dotted black line: non-cavity overall asymmetric stretching band of $N_2O$)

## 6. Hopfield coefficients

The Hopfield coefficients for LP and UP are shown in figure S6. UP contains 50-50 content of photon and R-branch when cavity mode is at ON-resonance with R branch maximum at ~8º, and it contains almost negligible contribution from P branch. Similarly, in case of LP, when cavity mode is at ON-resonance with P branch maximum ~1.5º, then LP consists of 50-50 content of photon and P branch with almost negligible contribution from R branch. And, for higher angles (i.e., higher inplane momentum), photon content in UP increases and P branch content in LP increase.

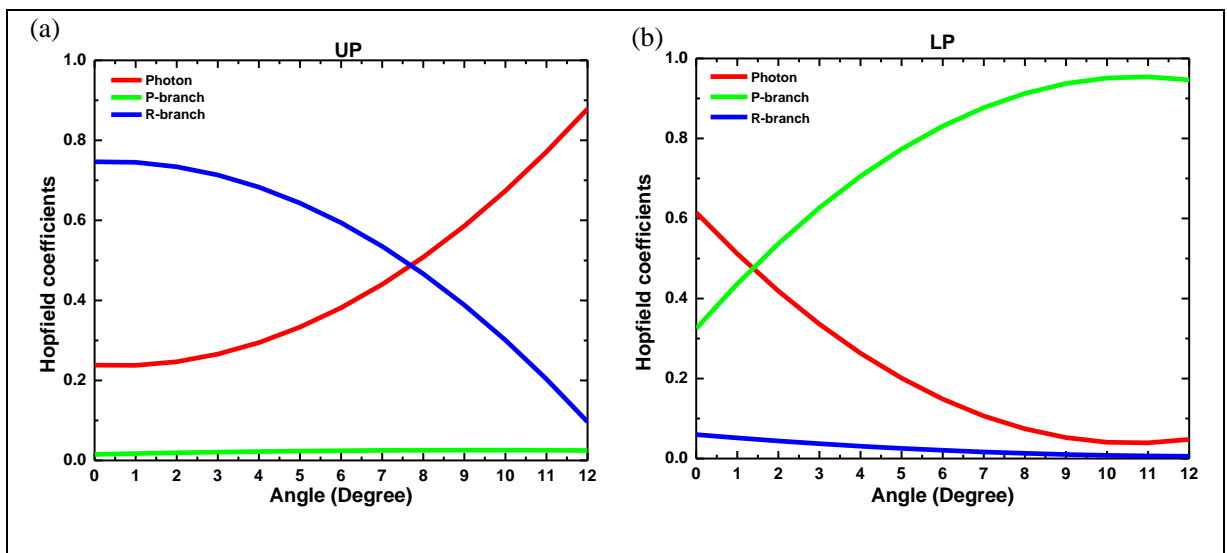

**Figure S6.** Hopfield fractions for **(a)** UP, and **(b)** LP with red, green and blue lines representing photon, P-branch and R-branch fractions, respectively.